\documentclass[sigconf,nonacm]{acmart}

\usepackage{amsmath,amsfonts}
\usepackage{algorithmic}
\usepackage{textcomp}
\usepackage{hyperref}
\usepackage{subcaption}
\usepackage{graphicx}
\usepackage{caption}
\usepackage{xcolor}
\def\BibTeX{{\rm B\kern-.05em{\sc i\kern-.025em b}\kern-.08em
		T\kern-.1667em\lower.7ex\hbox{E}\kern-.125emX}}
\usepackage{tikz}
\newcommand*\circled[1]{\tikz[baseline=(char.base)]{
		\node[shape=circle,draw,inner sep=1pt] (char) {#1};}}
\usepackage{colortbl}
\usepackage{multirow}

\usepackage{enumitem}
\usepackage{listings}
\usepackage{color}
\definecolor{dkgreen}{rgb}{0,0.6,0}
\definecolor{gray}{rgb}{0.5,0.5,0.5}
\definecolor{mauve}{rgb}{0.58,0,0.82}
\newcommand{\tab}[1]{Table~\ref{#1}}
\newcommand{\fig}[1]{Fig.~\ref{#1}}

\newcommand{\eg}{\textit{e.g.}, } 
\newcommand{\ie}{\textit{i.e.}, }

\newcommand{\etal}{\textit{et al.}, }

\newcommand{\flockmtl}{\textsc{FlockMTL}}
\newcommand{\duckdb}{\textsc{DuckDB}}

\begin{document}

\title[Beyond Quacking: Deep Integration of LMs and RAG into DuckDB]{Beyond Quacking: Deep Integration of Language Models and RAG into DuckDB}

\author{Anas Dorbani}
\affiliation{%
	\institution{Polytechnique Montréal}
}
\email{anas.dorbani@polymtl.ca}

\author{Sunny Yasser}
\affiliation{%
	\institution{Polytechnique Montréal}
}
\email{sunny.yasser@polymtl.ca}

\author{Jimmy Lin}
\affiliation{%
	\institution{University of Waterloo}
}
\email{jimmylin@uwaterloo.ca}

\author{Amine Mhedhbi}
\affiliation{%
	\institution{Polytechnique Montréal}
}
\email{amine.mhedhbi@polymtl.ca}

\begin{abstract}
Knowledge-intensive analytical applications retrieve context from both structured tabular data and unstructured, text-free documents for effective decision-making. 
Large language models (LLMs) have made it significantly easier to prototype such retrieval and reasoning data pipelines. 
However, implementing these pipelines efficiently still demands significant effort and has several challenges.  
This often involves orchestrating heterogeneous data systems, managing data movement, and handling low-level implementation details, \eg LLM context management.  
	
To address these challenges, we introduce \flockmtl: 
an extension for DBMSs that deeply integrates LLM capabilities and retrieval-augmented generation (RAG). 
\flockmtl~includes model-driven scalar and aggregate functions, enabling chained predictions through tuple-level mappings and reductions. 
Drawing inspiration from the relational model, 
\flockmtl~incorporates: 
(i) cost-based optimizations, which seamlessly apply techniques such as batching and caching; and 
(ii) resource independence, enabled through novel SQL DDL abstractions: \texttt{PROMPT} and \texttt{MODEL}, introduced as first-class schema objects alongside \texttt{TABLE}. 
\flockmtl~streamlines the development of knowledge-intensive analytical application and 
its optimizations ease the implementation burden. 
\end{abstract}

\maketitle

\section{INTRODUCTION}
\label{sec:introduction}

\textbf{Complexity of Workflows.} A variety of workflows are in the form of knowledge-intensive analytical applications, \ie they rely on integrating relevant context from structured and unstructured datasets to support data-driven decision-making. 
They further rely on analytics, 
semantic analysis, or a combination of both to take effective action. 
For example, consider an investigative analyst reporting on inquiries regarding a new vessel offence. The analyst might:
i)~consult tabular data to obtain specific vessel details;
ii)~interpret legal documents to assess the severity of the reported offence;
iii) aggregate the vessel's history with similar prior offences; and
iv)~identify and rank potential interventions. 

\noindent\textbf{Novel Data Pipelines.} The advent of LLMs has led to a technological step change. 
Their commoditization 
since the release of GPT-3~\cite{DBLP:conf/nips/BrownMRSKDNSSAA20} made it possible 
to implement 
pipelines that interleave: 
(i)~querying of tables; 
(ii)~retrieval 
of relevant tuples; and 
(iii)~generation and reasoning using LLM-predictions. 
These pipelines use disparate systems, 
\eg DBMSs and search engines, follow the RAG approach~\cite{DBLP:conf/nips/LewisPPPKGKLYR020} and use possibly tool calling~\cite{patil2023gorilla}.

\noindent\textbf{Implementation Challenges.} 
The development of such pipelines is reminiscent of the early data management era prior to the relational model~\cite{DBLP:journals/cacm/Codd70}. 
Data engineers 
currently make many low-level execution decisions, \eg choosing specific models for predictions, adapting prompts, 
managing LLM context, caching of 
predictions for reuse, and incorporating novel optimizations as they are released publically, and deciding when to use them. 
To make matters worse, any changes to the application requirements in terms of expected quality, latency, dollar cost, or scale require a major re-implementation. 
Beyond these execution decisions, 
the use of disparate systems results in significant data shuffling and missed co-optimization opportunities. 
This often pushes users to rely on 
DBMSs for initial simple querying and on re-implementing more complex operations within an orchestration layer. 

\renewcommand{\arraystretch}{1.2}
\begin{table*}[t!]
	\centering
	\begin{tabular}{lll}
		\rowcolor[HTML]{F1F1F7}\hline
		\multicolumn{3}{c}{\textbf{Scalar Functions: Map an input tuple to an output value using chat completion API or embedding API.}}\\ \hline
		\multirow{3}{*}{\textbf{Generic}}
		& \textsc{llm\_complete\{\_json?\}} & 
		uses an LLM and a user prompt to generate \emph{text} or structured \emph{JSON} output from an input tuple.\\
		& \textsc{llm\_embedding} &
		uses an LLM to generate an embedding vector (fixed-length array) from an input text value.\\
		& \textsc{fusion} &
		fuses $N$ scores from $N$ retrievers --- choices: \texttt{rrf} / \texttt{combanz} / \texttt{combmed} / \texttt{combmnz} / \texttt{combsum}.\\\hline
		\multirow{1}{*}{\textbf{Specialized}}
		& \textsc{llm\_filter} &
		uses an LLM and a prompt to return \emph{True/False} given an input tuple.\\\hline
		\rowcolor[HTML]{F1F1F7}\hline
		\multicolumn{3}{c}{\textbf{Aggregate Functions: Reduce multiple input tuples to a single output value using the chat completion API.}}\\ \hline
		\multirow{1}{*}{\textbf{Generic}} & \textsc{llm\_reduce\{\_json?\}} &
		uses an LLM and a prompt to generate \emph{text} or structured \emph{JSON} output from multiple input tuples.\\\hline
		\multirow{2}{*}{\textbf{Specialized}}
		& \textsc{llm\_rerank} &
		uses an LLM and a prompt to rank input tuples based on relevance.\\
		& \textsc{llm\_first/last} &
		uses an LLM and a prompt to return the most or least relevant tuple from multiple input tuples.\\\hline
	\end{tabular}
	\caption{Summary of the scalar and aggregate functions supported by \flockmtl.}
	\label{tab:summary-functions}
	\vspace*{-2em}
\end{table*}

\noindent\textbf{Our Approach.} We propose \flockmtl, an open-source extension for \duckdb~\cite{DBLP:conf/sigmod/RaasveldtM19} that enables the use of LLMs with scalar and aggregate functions. 
Using SQL's common table expressions (CTEs) with these functions leads to powerful pipelines that can interleave analytics with LLM-chained predictions.
Following in the tradition of declarative relational model, \flockmtl~uses cost-based optimization to alleviate the burden of implementing low-level execution details from developers and non-expert users. 

\noindent \textbf{Core Insights.} Our insights on important features designing and implementing \flockmtl~can be summarized as follows:
\begin{itemize}[leftmargin=*]
	\setlength{\itemsep}{2pt}
	\setlength{\parskip}{0pt}
	\setlength{\parsep}{0pt}
	\item  \emph{Flexible paradigm}: \flockmtl~supports a broad range of semantic operations, including classification, summarization, and reranking, through the use of  
	scalar and aggregate functions, all of which are summarized in \tab{tab:summary-functions}). 
	Additionally, \flockmtl~introduces some specialized built-in functions, \eg 
	reranking (first / last) and fusion (rrf, combanz, combmed, combmnz, combsum) that enable full hybrid search. 
	\item \emph{Resource-independence}: 
	Functions accept both model and prompt specifications as inputs. 
	\flockmtl~introduces two new DDL resource types: \texttt{MODEL}s and \texttt{PROMPT}s, treated as first-class schema objects akin to \texttt{TABLE}s. 
    This abstraction allows SQL queries to remain fixed while enabling model and prompt updates administratively, without requiring changes to application logic. 
	\item \emph{Seamless optimizations}: 
	Lower-level implementation like LLM context management, batching on input tuples in a single LLM prediction, caching and reusing of results, and predictions on unique, \ie deduplicated input values are handled seamlessly by \flockmtl.
	This reduces the complexity of integrating semantic operations and allows developers and data engineers alike to focus on higher-level application logic.
\end{itemize}

\newpage

\section{SYSTEM OVERVIEW}
\label{sec:overview}

LLMs and prompt engineering align well with SQL’s original goal of making data querying accessible to non-experts. 
\flockmtl's core feature is introducing semantic operations and hybrid search within SQL. 
Given that the majority of enterprise data is stored in RDBMSs, 
\flockmtl~relies on \duckdb~\cite{DBLP:conf/sigmod/RaasveldtM19}'s extension module. 

\duckdb~implements a state-of-the-art analytics engine and makes the DBMS internals easily extensible. 
Its extension module allows changes to SQL parsing, to the optimizer and execution engine, 
as well as the addition of new data types. 
\duckdb~has already a rich ecosystem of extensions that can be complementary. 
For instance, its extensions enable the querying of file formats such as \emph{Parquet} and \emph{CSV} as well as attaching to DBMSs such as \texttt{PostgreSQL} and \texttt{MySQL}. As such, users can write federated queries over multiple data formats and databases. 
As a DBMS of choice, the capabilities we add within \flockmtl~become instantly available across a variety of file formats and databases. 
\flockmtl~also provides an \texttt{ASK} functionality to turn a natural language queries into SQL augmented with \flockmtl's functions. 
It is easy to \texttt{INSTALL} and \texttt{LOAD} \flockmtl~as a community extension using:\vspace*{-0.1em}
\renewcommand{\lstlistingname}{Query}
\begin{lstlisting}[escapechar=\%,language=SQL]
	INSTALL flockmtl FROM community; LOAD flockmtl;
\end{lstlisting}

In the remainder of this section, we provide an overview of \flockmtl's new schema object resources (\texttt{MODEL} and \texttt{PROMPT}), functions, and optimizations. To illustrate resources and functions, we consider a simple use case involving the following table:\vspace*{-0.4em}
\begin{center}
	\texttt{research\_papers:} (\texttt{\underline{id},} \texttt{title,} \texttt{abstract,} \texttt{content})
\end{center}\vspace*{-0.4em}
In this scenario, the user is a researcher aiming to identify relevant papers, extract key insights, and generate summaries using SQL.

\subsection{Models and Prompts}
\label{sec:models-and-prompts}

Users can define reusable resources in the form of \texttt{MODEL}s and \texttt{PROMPT}s. These resources can be scoped to the current database using the \texttt{Local} setting, which is the default, or configured as \texttt{Global}, \ie accessible across all databases on a given machine.

\emph{Query 1} below shows how to define a global model named \emph{model-relevance-check}, configured to point to \texttt{GPT-4o-mini} with the server provider OpenAI. It also includes a local user-defined prompt for identifying papers relevant to join algorithms.

\begin{lstlisting}[caption={Prompt and Model Definitions},escapechar=\%,language=SQL]	
-- %\circled{1}% Define a model to use
CREATE GLOBAL MODEL('model-relevance-check', 'gpt-4o-mini', 'openai')
-- %\circled{2}% Define a prompt to check if the paper is a join algorithm
CREATE PROMPT('joins-prompt', 'is related to join algos given abstract')
\end{lstlisting}

Users have the flexibility to modify or delete these resources. 
When a resource is modified, \flockmtl~automatically creates a new version. Previous versions remain available for inspection and use, while the most recent version is applied by default unless specified otherwise.

\subsection{Functions}
\label{sec:functionality}

\flockmtl~enables semantic analysis through a combination of generic and specialized functions. 
The generic functions allow users to define operations that map or reduce tuples to text or \emph{JSON} outputs using LLMs available through Azure and OpenAI cloud services or locally through Ollama. These are summarized in \tab{tab:summary-functions}. 

Building on these generic functions, \flockmtl~also provides specialized functions that wrap the generic ones for common use cases. 
For example, \texttt{llm\_filter} converts LLM predictions into boolean values, and \texttt{llm\_first}/\texttt{llm\_last} use the output of \texttt{llm\_rerank} to select the most or least relevant tuple based on a ranking.

\noindent \textbf{Scalar Functions.} We showcase in the next query an example of semantic filtering, summarization, and extraction using \flockmtl's scalar functions. 
A scalar function maps each input tuple to a new attribute. 
\emph{Query 2} demonstrates this by identifying papers relevant to join algorithms. Then, for each relevant paper, it extracts keywords, determines whether the paper is empirical or theoretical, and summarizes its abstract in a sentence. The query uses three semantic operations: \texttt{llm\_filter}, which produces a boolean value; and \texttt{llm\_complete} and \texttt{llm\_complete\_json}, which return a string and a JSON object, respectively, via chat completion APIs.

\begin{figure*}[t!]
	\centering
	\includegraphics[width=.8\textwidth]{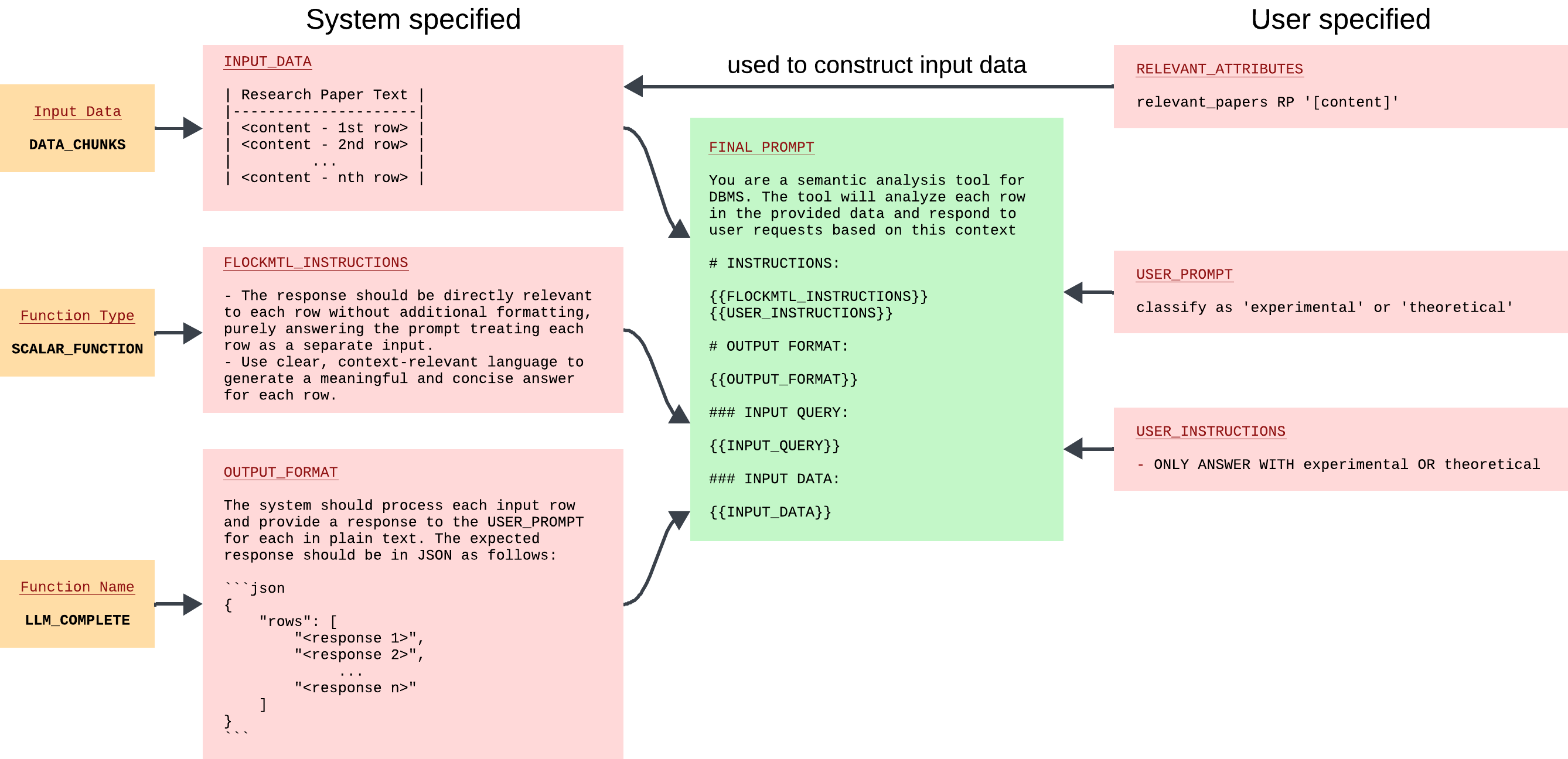}
	\vspace*{-1em}
	\caption{Prompt construction example for \emph{llm\_complete}.}
	\label{fig:prompt}
	\vspace*{-1.4em}
\end{figure*}

\begin{lstlisting}[caption={Finding relevant join algo papers.},escapechar=\%,language=SQL]
WITH 
-- %\circled{1}% Select papers related to join algorithms
relevant_xpapers AS (
	SELECT id, title, abstract, content
	  FROM research_papers P
	 WHERE llm_filter({'model_name': 'model-relevance-check'},
						 	  {'prompt_name': 'joins-prompt'},
						 	  {'title': P.title, 'abstract': P.abstract})
),
-- %\circled{2}% Summarize the paper's abstract
summarized_Papers AS (
	SELECT RP.id, RP.title, llm_complete({'model': 'gpt-4o'},
										{'prompt': 'Summarize the abstract in 1 sentence'},
										{'abstract': RP.abstract}) AS summarized_abstract,
	llm_complete_json({'model': 'gpt-4o'},
			{'prompt':'Based on the provided paper title, abstract, and content extract the following as JSON: {
							"keywords": <three relevant keywords>,
							"type": <specify if empirical or theoretical> }'},
			{'title': p.title, 'abstract': p.abstract})
	  FROM relevant_papers RP
)
SELECT * FROM summarized_Papers
\end{lstlisting}

More specifically, \texttt{llm\_filter} filters out non-relevant papers by applying a scalar function to each input tuple using the predefined \texttt{model-relevance-check} and an associated prompt. The resulting subset is then passed to \texttt{llm\_complete}, which summarizes each abstract using \texttt{GPT-4o}, and to \texttt{llm\_complete\_json}, which extracts keywords and the paper type in structured JSON form. While the use of CTEs in Query 2 is not required, it illustrates how \flockmtl~supports the chaining of LLM predictions through composable functions. Notably, only \texttt{llm\_filter} relies on a predefined model and prompt resources; the remaining functions specify these parameters directly within the query.

\noindent \textbf{Aggregate Functions and Full Hybrid Search.} 
We showcase in the next query an example of a full hybrid search pipeline within \flockmtl. 
To our knowledge, this is the first such implementation within SQL.  
\emph{Query 3} aims to find passages from research papers relevant to \emph{join algorithms in databases}.  
Among those, it further reranks the results to prioritize passages related specifically to \emph{cyclic join queries}. 
We consider a table containing previously extracted passages from publications: 
\texttt{research\_passages:} (\texttt{\underline{idx},} \texttt{content}).

The query breaks down the hybrid search into distinct steps. 
First, it computes the embedding for the user intent \emph{join algorithms in databases} using \flockmtl's \texttt{llm\_embedding}. It then performs a vector similarity scan, selecting the top $100$ most relevant passages from research. While this example could alternatively use the Vector Search Extension\footnote{\url{https://duckdb.org/docs/extensions/vss.html}}, the intention is to highlight the flexibility and direct use of \texttt{llm\_embedding}.
It also uses the Full-Text Search extension\footnote{\url{https://duckdb.org/docs/extensions/full_text_search.html}} to retrieve the top $100$ passages based on BM25 scores. 
The two result sets are then fused using a \texttt{FULL OUTER JOIN} and 
are normalized in this case using the max score for each retriever. 
Following fusion, the top-10 candidate passages are reranked using an LLM-based list-wise aggregation.  
This is achieved through \texttt{llm\_rerank}, a specialized function built on top of the generic aggregate interface.  
It applies a learned reranking model to the candidate list, inspired by the approach of Xueguang Ma \etal~\cite{DBLP:journals/corr/abs-2305-02156}, to assess their relevance to cyclic join queries.

\begin{lstlisting}[caption={Hybrid search to find top 10 passages on cyclic joins },escapechar=\%,language=SQL]
-- %\circled{1}% Compute the embedding for the input query
WITH Query AS (
	SELECT llm_embedding({'model':'text-embedding-3-small'}, {'query': 'join algorithms in databases'})::DOUBLE[1536] AS embedding), 
-- %\circled{2}% Scan vectors for similar search based on array_distance
VS AS (
   SELECT idx, content, array_cosine_similarity(Query.embedding, llm_embedding({'model':'text-embedding-3-small'},
    	                    {'content': content})::DOUBLE[1536]) AS score
     FROM research_passages
    ORDER BY vs_score DESC
    LIMIT 100)
-- %\circled{3}% BM25 retriever over chunked text contents of papers 
BM25 AS (
   SELECT idx, content, fts_main_research_chunks.match_bm25(index_column,
	'join algorithms in databases', fields:='content') AS score
     FROM research_passages
    ORDER BY bm25_score DESC
    LIMIT 100), 
-- %\circled{4}% Combine chunks with a fusion algorithm assuming the same score scale
top_10 AS (
	SELECT bm.content AS doc1, vs.content AS doc2
	  FROM BM25 FULL OUTER JOIN VS ON BM25.idx = VS.idx 
	 ORDER BY fusion(BM25.score::DOUBLE / (MAX(BM25.score) OVER ()), 
	                    VS.score::DOUBLE / (MAX(VS.score) OVER ()))
     LIMIT 10)
-- %\circled{5}% Rerank top 10 elements if they are relevant to cyclic join queries
	SELECT llm_rerank({'model':'gpt-4o'},{'prompt':'mentions cyclic joins'},
                       {'doc1': doc1, 'doc2': doc2})
	  FROM top_10;
\end{lstlisting}\vspace{-2.5em}

\begin{figure*}
	\centering
	\begin{subfigure}[b]{0.3\textwidth}
		\centering
		\includegraphics[scale=0.1565]{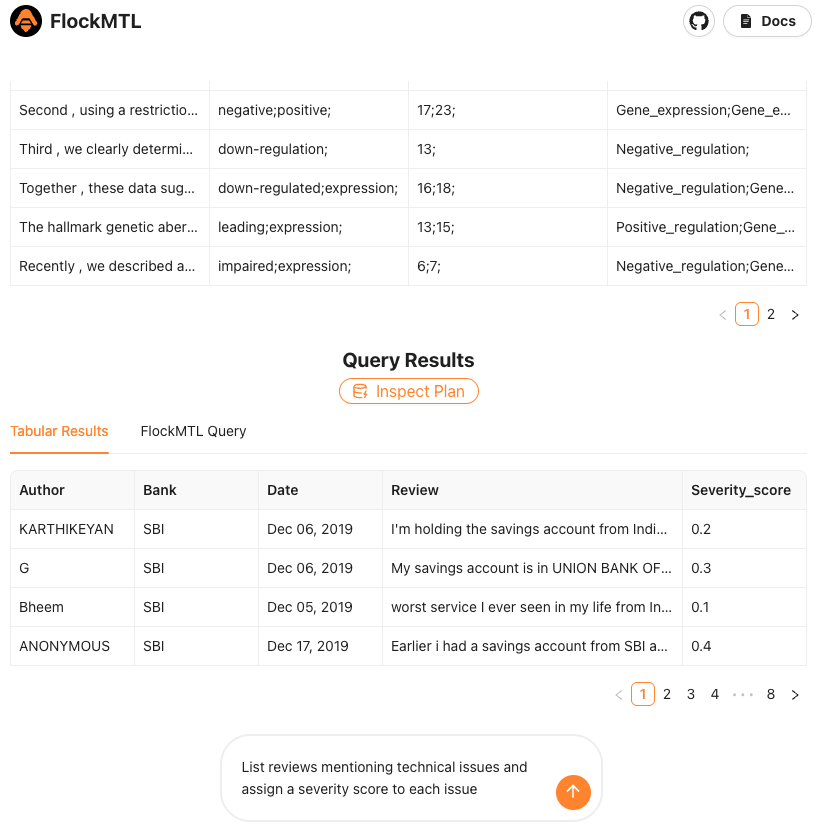}
		\caption{Data application with NL interface.}
		\label{fig:sub1}
	\end{subfigure}
	\begin{subfigure}[b]{0.68\textwidth}
		\centering
		\includegraphics[scale=0.28]{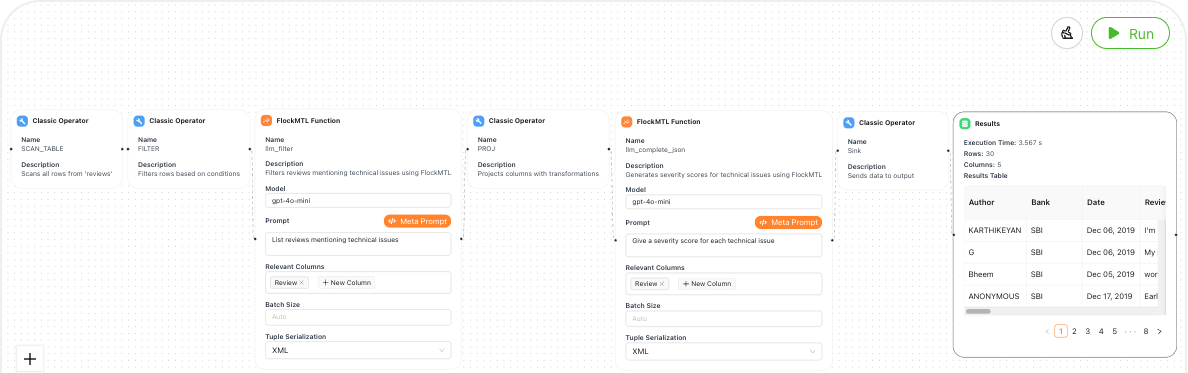}
		\caption{The plan inspection interface.}
		\label{fig:sub2}
	\end{subfigure}
	\caption{Screens of Prepared Demonstration for Users to Get Started.}
	\label{fig:test}
\end{figure*}

\subsection{Optimizations}
\label{sec:optimization}

\flockmtl~introduces several key optimizations to improve efficiency and usability: 
(i) meta-prompting for robust predictions and simpler user queries; 
(ii) batching tuples into a single request to improve latency;  
(iii) caching predictions for reuse within and across queries; and  
(iv) predicting over deduplicated values to avoid redundancy.  
Due to space limitations, we focus on (i) and (ii). 

\noindent\textbf{Meta-prompt.}  
In \flockmtl, users provide prompts intended for a single tuple (in the case of map functions) or for multiple tuples (in reduce functions). The system then composes a full prompt using the structured meta-prompt template shown in~\fig{fig:prompt}. 
This meta-prompt includes the user-specified content and augments it with formatting instructions, output expectations, and serialization of tabular input tuples. It is implemented to be KV-cache friendly.

\noindent\textbf{Batching.}  
When using \flockmtl's scalar functions, users write prompts for a single tuple. However, making an API call per tuple is inefficient, so \flockmtl~automatically applies batching to optimize inference. 
The system dynamically determines the batch size based on the input attribute values and the LLM’s context window size. It fills the prompt with as many tuples as possible until the context limit is reached, then sends a single batched request. 
If the LLM returns an error due to the output exceeding the context window, \flockmtl~automatically reduces the batch size by 10\% iteratively until a successful prediction is obtained. If a single tuple exceeds the output context size, the result is set to \texttt{NULL}.

On the \href{https://www.kaggle.com/datasets/dhavalrupapara/banks-customer-reviews-dataset}{Kaggle Bank Review dataset}, batching on a table scan query with a single scalar \flockmtl~function yields significant performance gains--achieving up to 7× speedup for chat completion map functions and 48× for embedding-based functions.

\section{Demonstration}
\label{sec:demo-scenarios}

\noindent\textbf{Goal.}  
Our goals with \flockmtl~is two fold.
First, we aim to showcase how easily users can build data applications using the \texttt{ASK} functionality, combining analytics and semantic analysis within an embedded RDBMS without the need to orchestrate multiple external systems. Second, we aim to highlight the importance of \flockmtl’s low-level optimizations by involving the audience in an interactive challenge. We added to our repo on github\footnote{\href{https://github.com/dsg-polymtl/flockmtl/}{https://github.com/dsg-polymtl/flockmtl/}} a demonstration for users to get started. 

\noindent\textbf{Interaction.}  
The landing page of the demonstration presents a data application where users can explore and interact with multiple tabular datasets sourced from Kaggle and spanning domains such as biomedical, academic, and product reviews. 
Users begin by viewing a preview of the dataset made of one or more tables. 
To explore the dataset, attendees can issue a natural language query, \eg ``\texttt{list reviews mentioning technical issues and assign a severity score to each issue}'' as done in \fig{fig:sub1} on a banking services review dataset, and \flockmtl's \texttt{ASK} functionality automatically generates a SQL query augmented with \flockmtl's functions. 
Users can inspect the generated SQL query and its output. 
This part of the demonstration illustrates the power of integrating semantic operations directly into SQL and its ease of use. 

Following this, by clicking on \texttt{Inspect} \texttt{Plan} for the generated query shown in \fig{fig:sub1}, users are taken to a separate interface for plan debugging and analysis. 
The separate interface shows the plan of our example query in \fig{fig:sub2}.
This query plan includes: 
(i) standard SQL operations such as scans and filters, and 
(ii) \flockmtl~specific functions, such as \texttt{llm\_filter} as well as \texttt{llm\_complete\_json}. 
The \flockmtl~function box on the UI contains additional system-level details, such as access to the full meta-prompt used, the serialization format, and the batch size chosen automatically by the system.

Users are first presented with the default setting where batch size is set to \texttt{Auto}, hiding the one \flockmtl~used. They can change it manually and select a different batch size that might match the system's performance and accuracy. 
The default serialization format shown by default is \texttt{XML}, but users may modify it to \texttt{JSON} or \texttt{Markdown}.
For instance, if a user sets the batch size to $30$ and reruns the query, they might observe a latency increase. 
We believe that the plan inspection highlights trade-offs in latency and prediction accuracy with different parameters. 
Finally, users can replace the full prompt using a Jinja template, and then compare it in both structure and output with the full one generated by \flockmtl. 

We believe that this hands-on demonstration shows both the accessibility of building semantic and analytical data applications in SQL when compared with alternative  systems~\cite{DBLP:conf/cidr/LiuRCCCCFKMSG25,DBLP:conf/cidr/BiswalPJKLGGZ25} and the value of \flockmtl's optimizations in reducing developer burden.\vspace*{0.85em}

\balance
\bibliographystyle{abbrv}
\bibliography{references}

\begin{thebibliography}{1}

\bibitem{DBLP:conf/cidr/BiswalPJKLGGZ25}
A.~Biswal and et~al.
\newblock Text2sql is not enough: Unifying ai and databases with tag.
\newblock In {\em CIDR}, 2025.

\bibitem{DBLP:conf/nips/BrownMRSKDNSSAA20}
T.~B. Brown and et~al.
\newblock Language models are few-shot learners.
\newblock {\em NeurIPS}, 2020.

\bibitem{DBLP:journals/cacm/Codd70}
E.~F. Codd.
\newblock A relational model of data for large shared data banks.
\newblock {\em CACM}, 1970.

\bibitem{patil2023gorilla}
S.~G.~P. et~al.
\newblock Gorilla: Large language model connected with massive apis.
\newblock {\em CoRR}, abs/2305.15334, 2023.

\bibitem{DBLP:conf/nips/LewisPPPKGKLYR020}
P.~S.~H. Lewis and et~al.
\newblock Retrieval-augmented generation for knowledge-intensive {NLP} tasks.
\newblock In {\em NeurIPS}, 2020.

\bibitem{DBLP:conf/cidr/LiuRCCCCFKMSG25}
C.~Liu and et~al.
\newblock Palimpzest: Optimizing ai-powered analytics with declarative query processing.
\newblock In {\em CIDR}, 2025.

\bibitem{DBLP:journals/corr/abs-2305-02156}
X.~Ma and et~al.
\newblock Zero-shot listwise document reranking with a large language model.
\newblock {\em CoRR}, abs/2305.02156, 2023.

\bibitem{DBLP:conf/sigmod/RaasveldtM19}
M.~Raasveldt and H.~M{\"{u}}hleisen.
\newblock Duckdb: an embeddable analytical database.
\newblock {\em SIGMOD}, 2019.

\end{thebibliography}

\end{document}